\documentclass[MSNbibl,number,citesort,dvips]{arxstspdf}
\usepackage{dcolumn}
\usepackage{graphicx}
\usepackage{flushend}
\usepackage{stfloats}

\volume{26}
\issue{4}
\pubyear{2011}
\firstpage{528}
\lastpage{542}
\doi{10.1214/10-STS350}

\makeatletter
\newcolumntype{d}[1]{D{.}{.}{#1}}
\makeatother

\begin{document}
\begin{frontmatter}

\title{2004 Venezuelan Presidential Recall Referendum (2004 PRR):
A Statistical Analysis from the Point of View of Electronic Voting Data Transmissions}

\runtitle{2004 Venezuelan Presidential Recall Referendum (2004 PRR)}

\begin{aug}
\author[a]{\fnms{Isbelia} \snm{Mart\'{i}n}\corref{}\ead[label=e1]{isbeliam@usb.ve}}

\runauthor{I. Mart\'{i}n}

\affiliation{Universidad Simon Bolivar}

\address[a]{Isbelia Mart\'{i}n is Professor, Department of Physics, Universidad Simon Bolivar, Caracas 1083,
Venezuela \printead{e1}.}

\end{aug}

% ABSTRACT
\begin{abstract}
Statistical comparisons of electoral variables are made between groups
of electronic voting machines and voting centers classified by types of
transmissions according to the volume of traffic in incoming and
outgoing data of machines from and toward the National Electoral
Council (CNE) totalizing servers. One unexpectedly finds two types of
behavior in wire telephony data transmissions and only one type where
cellular telephony is employed, contravening any reasonable electoral
normative. Differentiation in data transmissions arise when comparing
number of incoming and outgoing data bytes per machine against total
number of votes per machine reported  officially  by the CNE. The
respective distributions of electoral variables for each type of
transmission show that the groups classified by it do not correspond to
random sets of the electoral universe. In particular, the distributions
for the NO percentage of votes per machine differ statistically across
groups. The presidential elections of 1998, 2000 and the 2004
Presidential Recall Referendum (2004 PRR) are compared according to the
type of transmissions in 2004~PRR. Statistically, the difference between
the empirical distributions of the 2004 PRR NO results and the 2000
Chavez votes results by voting centers is not significant.
\end{abstract}

% KEYWORDS
\begin{keyword}
\kwd{Electronic voting}
\kwd{electoral data transmission}
\kwd{recall referendum}
\kwd{Venezuelan elections}.
\end{keyword}

\vspace*{-12pt}
\end{frontmatter}

%s1 ###
\section{Introduction}\label{s1}

During the Venezuelan Presidential Recall Referendum (PRR) held on
August 15 of 2004, voters used electronic voting machines to cast their
votes. A NO or YES vote meant a pro-government or anti-government vote
respectively. In order to investigate the\vadjust{\goodbreak} trustworthiness of the
electoral results, we carry out a forensic analysis of the government
official National Electoral Council (CNE) electoral results transmitted
by machines nationwide and of data contained in Remote Authentication
Dial-In User Service (RADIUS) logs of transmissions produced by
authentication, authorization and accounting (AAA) servers used in wire
and cellular transmissions between voting machines and totalizing
servers \cite{2,3,4}.

While in this paper we only explore transmission data collected
during the Venezuelan PRR of 2004, some of the methods presented here
can be applied in different contexts. In particular, the discussion in
the manuscript can inform governments and other international
organizations who wish to plan electoral audits in the future, in
Venezuela or elsewhere. Given the increasing\vadjust{\goodbreak} popularity of electronic
voting machines world-wide, the development of monitoring and auditing
methods to guarantee the reliability of electronic voting processes is
critically important.

Transmission data correspond only to communications through wire and
cellular (mobile) telephony but cover 98.05\% of the universe of
electronic voting machines used in the electoral event. Four
independent sources of information as far as transmissions through wire
and cellular telephony were used: Two of the sources correspond to
RADIUS logs, for wire and cellular transmissions respectively,
containing information on several technological variables for
individual voting machines, among them: amount of octets (bytes) of
incoming and outgoing data to CNE totalizing servers, start and stop
connection times to CNE totalizing servers, amount of packets of
incoming and outgoing data, identification of users, hosts and routers,
etc. The third and fourth sources are based on a report draft made by
the wire telephone company to the CNE and the actual automated tallies
printed by machines respectively. These reports served to cross-check
information with the RADIUS logs to offer validity to the same logs as
reliable sources of information as far as volume of  transmitted data,
connection times and duration of sessions by machine.

In the present article the forensic analysis consists first in studying
the behavior of machines according to volume of data transmitted and
received, and relating it to the vote totals counted electronically and
transmitted by each voting machine. Second, a complementary statistical
study is performed that puts emphasis only on the heterogeneity of the
behavior of groups of machines found in the first analysis that allowed
a classification according to transmissions to CNE totalizing servers.
Directionality of data transmission is not relevant in this part since
the statistical analysis is not affected by it.

According to electoral norms, all machines had to transmit  vote totals
scrutinized by the same machine. Also, the information transmitted must
include polling station code numbers, poll's closing time, number of
registered voters, number of votes and vote totals results. This
information had to be contained also in paper reports produced by the
machine. Furthermore, the amount of bytes needed to transmit the  data
should have been exactly the same for all machines in the country,
regardless of geographical location and any other differences such as
polling center's codes\vadjust{\goodbreak} or voting volume at the center. Also, the
software for recording votes, counting and transmitting results should
be the same for all machines employed. The electronic information on
tallies had a fixed length in bytes. Thus, any disparity in volume of
data transmitted, not accounted for ordinary transmission errors as
eventual lost packets of data, is unexpected given the electoral
standards. Even more surprising is to find a linear dependence of
transmitted data bytes  on individual ballots for a high percentage of
machines. This fact will be our concern. Furthermore, given the
electoral normative, when a call session is established between the
totalizing server and a machine at the closure of polls, only data
relating to authentication, authorization and acknowledgment of
reception of data should be sent to the machine. This amounts to
a~fixed volume of data, smaller in size than the one related to vote
results sent by the machine. But the findings contradict these
expectations.

In this analysis, the electronic voting machines were classified
according to the amount of data units in bytes that were sent from
machines to CNE totalizing servers (Outgoing Data) and according to the
amount of data received by voting machines from totalizing servers CNE
(Incoming Data). For this study only the data transmissions of the last
successful connection between machines and CNE totalizing servers were
taken into account. We suppose a priori that when several calls were
made from the same machine it was due to defective transmissions, that
is the reason why the last connection is presumed to be the successful
one and that the amount of data transmitted in both directions in that
occasion was the expected information according to the programmed
procedure. Machines communicating more than once amounted to less than
15\% of  the total.

We find that the electronic voting machines fall into three groups:
High Traffic (A) for wire transmission machines if the number of
outgoing bytes added to incoming bytes surpasses 23 thousand bytes, Low
Traffic (B) for wire transmission machines with total data traffic
lower than 7.5 thousand bytes and machines communicating via cellular
telephony (C). Differences on volume of data transmitted and received
are accompanied by differences in number of packets of data and causes
of termination of sessions. In fact, group A has transmissions with the
same number of packets being received and sent (symmetric transmission)
with call terminated by totalizing servers. Group~B has few packets
sent but many received (asymmetrical transmission) and calls terminated
by machines. For detailed information on network platforms, protocols
used and more see Malpica, Velasco and Martin~\cite{1}.

The voting centers, at the same time, were equally classified as High
Traffic centers (A), Low Traffic and Cellular ones, (B) and (C)
respectively, according to whether voting machines in a center fell
into the three groups mentioned above. In the case of existing mixed
wire and cellular transmissions in a~center, the classification  was
made according to the highest number of machines of a particular type~%
A, B or C, in general the number of mixed centers in each category A
and B is less than 10\% of the total.

There were no voting centers with mixed High (A) and Low (B) Traffic
machines. In general, voting centers could have from 1 to 18 machines
grouped in electoral tables, which could in turn accommodate from 1 to
3 voting machines. The typical voting center housed 4 machines. In the
Venezuelan electoral system, voting centers are arranged into parishes,
the latter into municipalities and several municipalities make a state.
Venezuela is divided into 24 states.

The rest of this manuscript is organized as follows. We first explore
the characteristics of the three groups A, B and C of voting machines
comparing the volume of data in bytes transmitted from and toward the
totalizing CNE servers relative to the number of votes cast in each
machine as reported by CNE.  Indeed, the classification on the basis of
empirical observation of differences in the pattern of graphs is
justified. The results from these exploratory analyses are presented in
Section \ref{s21}. We then investigate whether voting centers classified as
A, B or C exhibit different distributions of the following variables:
\begin{itemize}
\item The percentage of abstentions per machine nationwide.
\item The percentage of NO votes per machine nationwide.
\item The percentage of NO votes per voting center, compared to what was observed during the
presidential elections of 1998 and 2000.
\end{itemize}
All of these results are presented in Section \ref{s22}. Finally, some
brief conclusions are offered in Section~\ref{s3}.

%t1 ###
\begin{table*}[t]
\tablewidth=363pt
  \caption{}\label{t1}\vspace*{-12pt}
\begin{tabular}{@{}lcccc@{}}
\hline
&\textbf{High Traffic---wire}&\textbf{Low Traffic---wire}&\textbf{Cellular (C)}&\textbf{Total}\\
&\textbf{(A)}&\textbf{(B)}&&\\
\hline
Voting centers&1,876&1,573&972&4,421\\
Number of voting&&&&\\
\quad machines in centers&8,185\tabnoteref{a}&7,383\tabnoteref{a}&3,124\tabnoteref{b}&18,692\tabnoteref{c}\\
Number of machines&&&&\\
\quad in each class&7,535&6,702&4,455&18,692\tabnoteref{c}\\
Numbers of votes&3,695,415&3,300,896&1,357,733&8,354,044\tabnoteref{d}\\
\quad and \% of total&(43.44\%)&(38.80\%)&(15.96\%)&\\
\hline
\end{tabular}
\tabnotetext[*]{a}{Includes voting machines with cellular transmission.}
\tabnotetext[**]{b}{Includes voting machines with High Traffic transmission (0.5\%).}
\tabnotetext[\dagger]{c}{Represents 98.05\% of automated 2004 PRR.}
\tabnotetext[\dagger\dagger]{d}{Represents 98.20\% of automated 2004 PRR.}
\vspace*{-3pt}
\end{table*}

%f1 ###
\begin{figure*}[b]

\includegraphics{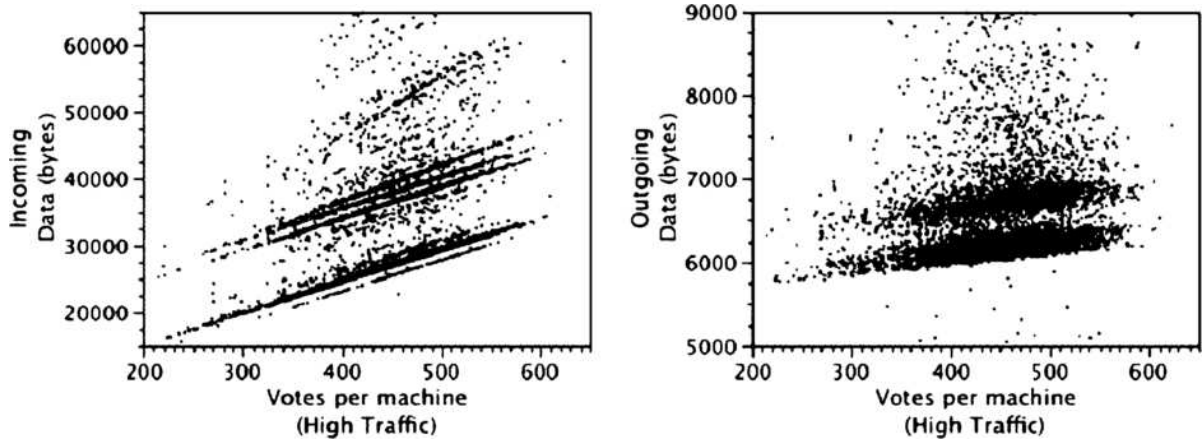}

\caption{Graphs of amount of bytes in data emitted and received by
each machine versus number of total votes per machine for the group of
High Traffic transmission. In the Outgoing data graph representing a
sample of 6,579 machines, it is possible to differentiate two subgroups
of machines related to two clusters: one superior cloud (G2 with 2,166
machines) and another inferior cloud (G1 with 4,413 machines).}\label{f1}
\end{figure*}

%s2 ###
\section{Results}\label{s2}

The electronic voting machines transmitted via wire, mobile and
satellite telephony. The machines transmitting via wire\vadjust{\goodbreak} telephony fall
into two groups, High Traffic (A) and Low Traffic (B), according to
whether the amount of data received plus the amount of data sent  is in
the range of 23,000 to 63,000 bytes for the High Traffic class and from
1,500 to 7,500 bytes in the Low Traffic class. When the electoral
variables by region are studied, it will be corroborated that the
classification of High and Low Traffic is bound to the telephone area
codes. We find whole municipalities in regional states whose voting
machines fall in one or another category. The group of machines that
transmitted via cellular is not much different to the wire High Traffic
group as far as the pattern of Bytes vs. Votes is concerned and the
volume of bytes transmitted but, technologically, they are not
comparable to the wire telephony. That is the reason why it is included
in this study as a separate group. In the present analysis machines
that have communicated via satellite are not mentioned for lack of
data.

The classification of High, Low Traffic and Cellular for voting centers
corresponds to those centers where most of their machines classified in
some of the mentioned classes. It is possible to find around 8\% to 9\%
of machines that communicated via cellular in some of the High and Low
Traffic centers; this could be justified by transmissions failures of
wire telephony or for a way to speed up the process of data
transmission when few wire lines were available. There are no centers
where machines transmitted in both High Traffic and Low Traffic groups.
The inclusion of cellular transmissions with High or Low Traffic wire
transmissions responds to the fact that analyzed electoral variables do
not vary significantly among machines in the same center.

Thus, 4,421 voting centers are grouped in 1,876 High Traffic centers
housing 8,185 voting machines including cellular machines, 1,573 Low
Traffic centers with 7,383 machines including cellular transmissions
and 972 Cellular centers having 3,124 machines with the exception of 17
machines that fall into the category of High Traffic wire telephony.
The total number of voting machines in this study is 18,692,
corresponding to 98.05\% of the 19,064 voting machines officially used
in the 2004 PRR and for which registries are known through electronic
tally reports. Table \ref{t1} shows the number of machines and centers with
transmission via wire and cellular telephony according to the volume of
traffic, and, also, the number of entered effective votes in each
category from a universe of 8,505,867 automated votes in the 2004 PRR
according to electronic tally reports.

%s2.1 ###
\subsection{Incoming and Outgoing Data versus Votes between
Electronic Voting Machines and CNE Totalizing Servers}\label{s21}

\begin{figure*}

\includegraphics{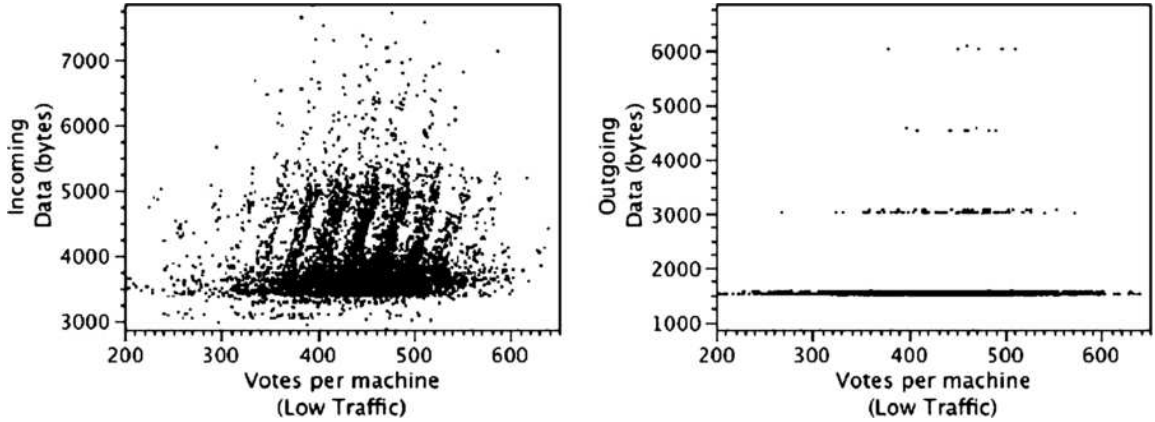}

\caption{Graphs of amount of bytes in data emitted and received by
each machine versus number of total votes by machine in 2004 PRR for
the group of machines with Low Traffic transmission.}\label{f2}
\vspace*{-3pt}
\end{figure*}

\begin{figure*}[b]
\vspace*{-3pt}
\includegraphics{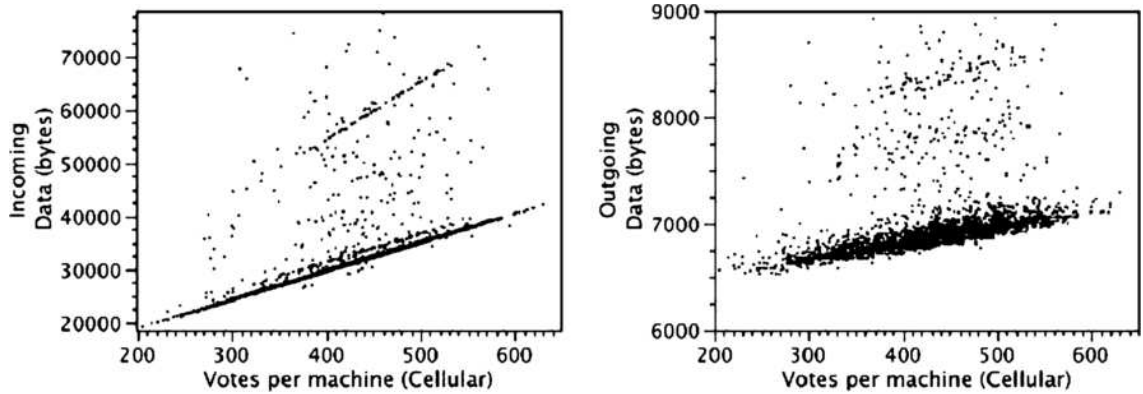}%
\caption{Graphs of amount of bytes in data emitted and received by
each machine versus number of total votes by machine in 2004 PRR for
the cellular transmission group of machines.}\label{f3}
\end{figure*}

For each group of machines, technological and electoral variables are
represented in an $x$--$y$ plane. The number of total votes by machine
reported by official  reports is in the $x$-axis and the amount of bytes
in the data that left and came into the machines during the
transmission is in the $y$-axis. The points on the plane represent
individual machines that reported a determined number of total votes,
and the data in bytes they emitted to transmit the voting results to
CNE totalizing servers (Outgoing data), as well as the data received by
the machine from CNE totalizing servers during the established sessions
of communication (Incoming data). (See Figures
\ref{f1}--\ref{f3}.)\looseness=1

%f2 ###

%f3 ###

A strong correlation between bytes in the incoming data being
transmitted and the amount of votes per machine is observed in High
Traffic and Cellular groups of machines (groups A and C). But the
behavior of the Low Traffic group of machines (group~B) is totally
different. A linear relation between number of votes and bytes in the
transmitted data may be observed only by a small number of machines
framed by a high dispersion of points.

Outgoing data transmissions, on the other hand, show clusters of points
with small correlation to the number of votes in groups A and C, but no
correlation exists for the horizontal plot on the Low Traffic group.

%s2.1.1 ###
\subsubsection{A---High Traffic transmissions}\label{s211}

Within the High Traffic group, it is possible to observe two clear\-ly
differentiated clusters in the Outgoing data graph, one  that we will
call the G1 subgroup and the other the G2 subgroup. These subgroups
correspond to points falling into the various parallel straight lines
that gather in the Incoming data graph. Subgroup G1 in the Outgoing
graph is related to the lower straight lines in the Incoming graph.

The perception of a greater dispersion of points shown in the Outgoing
data graph compared to the one in the Incoming data graph is due to
different scales involved in the volume of data sizes in both graphs.
Dispersion in graphs as well as various straight parallel lines in the
Incoming data graph may be related to retransmission of  packets of
data lost during transmission. One should expect that the higher the
number of packets of data to be transmitted the higher would be the
possibility of losing some of them during transmission, so they are
retransmitted and the number of bytes required for sending the same
information should increase. Since the number of bytes in packets would
differ, retransmission could\vadjust{\goodbreak} produce a random dispersion pattern. Also,
some dispersion could be  pointing to a mismatch between the number of
votes reported by the electronic machines and the actual number of
votes transmitted by machines to totalizing servers. This inference
relies on the presumption that any difference on data transmission
bytes among machines could only be related to the amount of votes being
reported since the rest of the information sent from machines had a
fixed amount of bytes assigned in the memory by the software according
to electoral norms. Parallel lines in graphs may also be produced when
more packets of data are transmitted intentionally.

%f4 ###
\begin{figure*}

\includegraphics{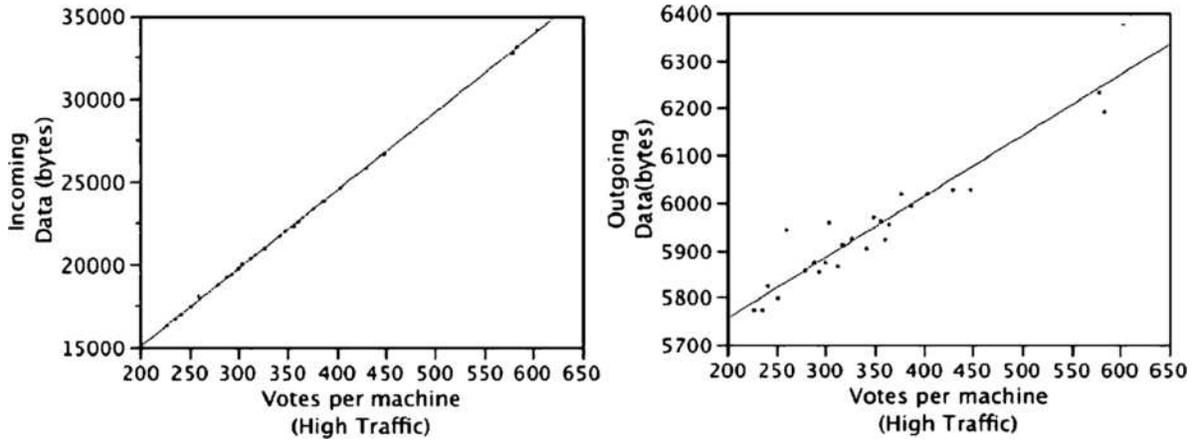}

\caption{Graphs of amount of bytes in data emitted and received by
each machine versus number of total votes per machine for a High
Traffic machines sample extracted from the lowest straight line shown
in Figure \protect\ref{f1}. The straight lines show lines of regression with a slope
of 47.11 bytes by vote for Incoming data with a 0.2\% error and 1.28
bytes by vote for Outgoing data with 6.0\% error in the same
machines.}\label{f4}
\end{figure*}

In order to determine a typical value of the relations between incoming
and outgoing bytes in High Traffic machines with the reported votes by
machine, a sample of machines that fall on the lowest straight line
with the largest number of points of the Incoming data graph of
Figure \ref{f1} was taken randomly. In the Outgoing data graph, these
machines are in subgroup G1. Then, regressions for Incoming data bytes
with respect to votes by machine as well as Outgoing data bytes against
votes by machine of the same selected machines were calculated. The
graphs for the selected sample regressions are in Figure \ref{f4}.
The linear regressions show a relation between bytes and votes given by
the following equations:
\begin{eqnarray*}
&&\mbox{Incoming data bytes}
\\
&&\quad= 5606\,( \pm\,52) + 47.11\,(
\pm\,0.14) \mbox{ Votes},
\\
&&\mbox{Outgoing data bytes}
\\
&&\quad= 5498\,( \pm\,30) + 1.28\,(
\pm\,0.08) \mbox{ Votes}.
\end{eqnarray*}

Segregating the High Traffic machines into the above mentioned
subgroups, corresponding to the superior cluster (G2) and the inferior
cluster (G1) in the Outgoing data graph, it is possible to corroborate
that the average of received data bytes by machines in the inferior
cloud (G1) is around 27,000 bytes and the one on the superior cloud is
of 37,000 bytes, whereas the average in the emitted data is around 6,200
bytes in the first case and of 6,700 bytes in the second one. It is
found that these two dissimilar behaviors are simultaneously occurring
in machines of the same electoral table in the same voting center for a
high number of voting centers. Of the 1,876 High Traffic centers
studied, 1,051 centers (56\%) correspond to the category of mixed
tables, 663 centers (35\%) with all machines in the inferior cloud and
162 (9\%) in the superior one. It is necessary to notice that the
majority of these machines only connected once with the totalizing
servers from which it is deduced that the connections were unique and
successful.

Proportions $56\dvtx  35\dvtx  9$ for centers with mixed subgroups, inferior and
superior subgroup machines respectively, may be considered as
originating from a~random sample of the universe of High Traffic
centers if the probability of occurrence of a machine with traffic in
the superior subgroup is 0.33 and for the inferior one is of 0.67 which
are the ratios shown by the subgroups to the universe of 6,579 machines
(2,166 in superior cloud and 4,413 in the inferior one).

It follows that machines located in the same electoral table using
presumably the same source code, the same telephone area codes,
sometimes the same telephone line and local networks with the same
technology, similar electoral populations and with similar physical
conditions behaved in such a different manner in both the reception and
emission of data, even though the relations of bytes to vote remained
more or less the same. The distributions of votes by machine in both
groups differ in average in 10 votes per machine, being greater in the
superior subgroup. Nevertheless, there are no statistical differences
in the average of percentage of YES and NO votes reported by automated
reports.

%f5 ###
\begin{figure*}[b]

\includegraphics{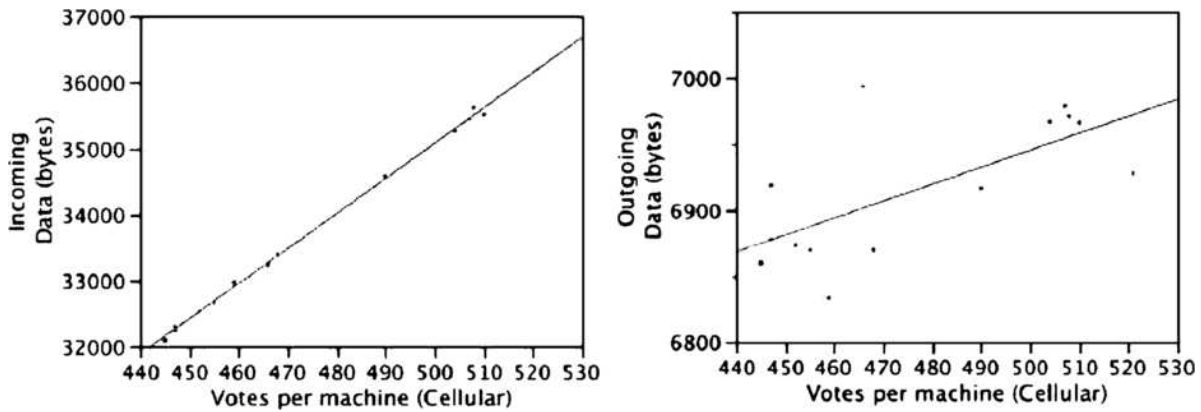}

\caption{Graphs of amount of bytes in data emitted and received by
each machine versus number of total votes per machine for 14 voting
machines transmitting via cellular telephony in the Colegio
Internacional de Caracas located in Baruta municipality. The straight
lines show lines of regression with a slope of 53.25 bytes by vote for
Incoming data with a 0.9\% error and 1.28 bytes by vote for Outgoing
data with 30\% error in the same machines.}\label{f5}
\end{figure*}

It is difficult to technologically explain a behavior so systematically
different in the emitted and recei\-ved data in High Traffic voting
machines located in the same electoral table of the same polling
center.\looseness=-1

%s2.1.2 ###
\subsubsection{B---Low Traffic transmissions}\label{s212}

In the  Low Traffic group in wire telephony  there is no relation
between the volume of Outgoing data and the number of votes computed in
each machine, suggesting the information transmitted was homogeneous
for the machines of this sector. Practically no dispersion is shown in
comparison with the behavior in the High Traffic group. This behavior
corresponds more to the expected one when only the information on vote
totals is transmitted and there are no packets retransmitted.

On the other hand, the Incoming data of voting machines show a regular
pattern that depends on the number of votes in the machines only in a
small sector. 27.5\% of the machines are in the vertical segments of
the graph related somehow to number of votes. But the rest of the
machines in the horizontal cluster do not show any relation between
Incoming data bytes to transmitted votes. This graph also shows a great
deal of dispersion. In general, machines in the same electoral table
could be located in any one of the two mentioned sectors.\looseness=1

The pattern of Incoming data versus votes in the Low Traffic machines
does not seem to respond to the model of individual vote transmission,
as it is the case for the High Traffic group. Nevertheless, those
machines whose Incoming data bytes are correlated with votes do so in a
nonhomogeneous way, the proportional relations between bytes and votes
go from~41 to 46 bytes per vote; these proportions are comparable in
magnitude to those observed in the High Traffic group but only among
machines that differ approximately in 30 votes.

Once again, it can not be technologically explained that machines in
the same electoral table have\vadjust{\goodbreak} behaviors so differentiated in bytes
transmissions; some of them are in the vertical segments of the graph
and other ones are in the cluster base.

%s2.1.3 ###
\subsubsection {C---Cellular transmissions}\label{s213}

In the Cellular transmissions machines (group C),  there
is a strong correlation between votes and Incoming data bytes
transmitted, much in the fashion of the High Traffic group. The same
may be said of the Outgoing data bytes against number of votes in each
machine. To illustrate behavior in this group (Figure \ref{f5}), a
particular voting center where transmissions from machines were all
through cellular telephony is chosen. This voting center was located in
a municipality where the majority of centers fell into the group of Low
Traffic transmissions.

The regressions are as follows:
\begin{eqnarray*}
&&\mbox{Incoming data bytes}
\\
&&\quad= 8461\,(\pm\,246) + 53.25\,(\pm\,0.51) \mbox{ Votes},
\\
&&\mbox{Outgoing data bytes}
\\
&&\quad= 6304\,(\pm\,188) + 1.28\,(\pm\,0.39) \mbox{ Votes}.
\end{eqnarray*}

It is observed that the slopes of straight lines correspond to 53 bytes
per vote for the Incoming data versus votes relations and of 1.28 bytes
per vote in the slope for Outgoing data versus votes with errors
indicating the degree of dispersion. The pattern shown indicates
transmission of individual votes as in the case of the High Traffic
group. Differences in the volume of data transmission compared to the
High Traffic wire telephony are due to differences in transmission
technology with data bytes measurements made at different levels.

%s2.1.4 ###
\subsubsection{General results in transmissions}\label{s214}

The preceding discussions suggest that either the programming of
electronic voting machines for data transmission or the programming in
the CNE totalizing servers to handle data transmissions to machines
behaved in different ways for two sets of groups of machines, groups A
and C compared to group~B. Although the transmissions through cellular
telephony would not be comparable with that of wire telephony because
of differences in technology, the remarkable differences in the volume
of data and patterns of transmission between the groups of High and Low
Traffic machines in wire telephony cannot be satisfactorily explained
under the electoral rules.

Electoral rules required that each machine should do the counting of
recorded votes and then its results be transmitted to the totalizing
servers. That is, tallies and not individual votes should be
transmitted. Transmission of tallies required a fixed amount of bytes
per machine, the same is true for the authorizing, acknowledgement
answers and transmission certificates sent from totalizing servers to
machines; in these cases horizontal straight lines should be expected
in Incoming and Outgoing data graphs against votes per machine, with
perhaps some variability in the number of bytes mainly for Outgoing
data.

Therefore, the dependence of the amount of data bytes on the number of
votes is inexplicable under the premise of vote totals transmissions,
which was supposed to be the electoral normative. In fact, graphs show
clearly a pattern for transmission of individual votes in both
directions to and from totalizing servers for the High Traffic and
Cellular groups. Also, if the programming software in the machines was
the same for all machines, one does not understand either the
differences in the types of linear relations with the number of votes
reported in every voting machine, or the volumes of data reported in
logs since the transmitted information must have equivalent sizes in
all wire telephony cases.

%f6 ###
\begin{figure*}[t]

\includegraphics{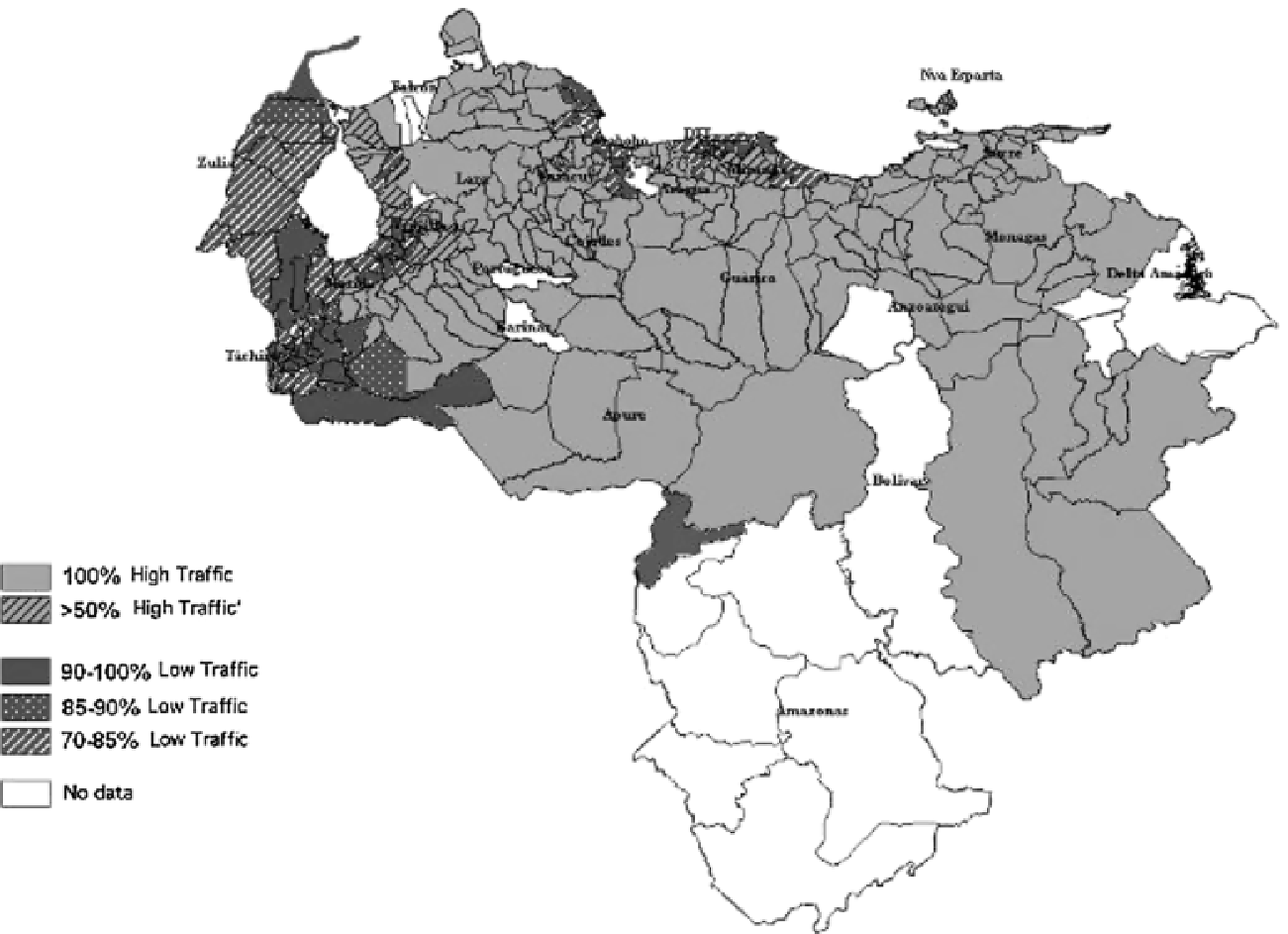}

\caption{\textup{Data transmission in municipalities and states}.
The map shows Venezuela divided by municipalities with some States
marked. Full light gray color municipalities include High Traffic wire
and Cellular transmissions. Striped gray color regions refer to
municipalities containing some parishes with Low Traffic transmissions.
The dark gray color municipalities are regions with a majority of Low
Traffic transmissions mixed with small percentages of Cellular
ones.}\label{f6}
\vspace*{-6pt}
\end{figure*}

On the other hand, a systematic behavior in the transmissions going
from machines to servers and also from servers to machines might
suggest a programmed intentionality.

Other findings are also consistent with the suggestion of intentional
tampering with the vote counting and transmission process. The
irregular distribution of groups of machines, mainly in wire telephony,
in different parishes and municipalities with no overlapping, cannot be
explained reasonably by random technological causes.\vadjust{\goodbreak} If the difference
in volumes of traffic in wire telephony is due to a  technological
variable, then it would be difficult to understand why the Aragua state
in its totality behaves technologically different to nearby states like
Carabobo and Miranda that share the same telephone network. The same
occurs between contiguous parishes in the same municipality in
Carabobo, Miranda, Merida and Trujillo. A map with occurrences of A, B
and C machine groups by municipalities is shown in Figure~\ref{f6}
(Data transmission in municipalities and states).

%s2.2 ###
\subsection{Empirical Distributions of Electoral Variables Across
the Three Groups of Voting Machines and Centers}\label{s22}

In what follows  differences and similarities between distributions of
several variables for the three groups of voting machines and centers
are studied statistically. The reason to carry out this additional
analysis is to shed light on the incidence of the differentiation in
the voting machines transmissions on electoral results or vice versa.

A priori we could infer that groups A, B and C of the machines would
show differences in electoral variables like abstention and vote
results because the number of urban voters is higher in group~B than in
other groups, as we could gather from the geographical distribution of
groups of machines. In fact, we should find that the three groups are
not random samples of an electoral universe.  But then if there was
intentional tampering with the votes, the grouping of the machines and
centers must be somehow associated with electoral results since there
seem to be no technological factors that can explain the groups. On the
other hand, if there was an innocent reason for loading two different
software packages in either machines or servers, the electoral
authorities should have mentioned this fact prior to the election.
Suspicions arise mainly because voting machines were connected to
totalizing servers at CNE headquarters, before tallies were printed by
machines locally. Even more, previously planned audits were not fully
carried out  and ballot boxes with paper tickets produced by the
machines were not allowed to be opened.

In this light, a question of interest is the following:  is the linear
dependence of Outgoing and Incoming data bytes on votes related to
virtual votes and tampering of electoral results? If that was the case
for groups A and C, what happened to group~B where the linear
dependence shows only for 27.5\% of the machines?\vadjust{\goodbreak}

If vote results were tampered with, it seems logical to think that the
made-up voting patterns were not made up during the electoral event
but, rather, were determined in advance of the election. If so, an
approach to generate plausible distributions of NO and YES votes across
the various voting centers would be to mimic what was observed during
the 1998 and 2000 presidential elections. Thus, we explore the
similarities and the differences between the electoral results reported
for the 2004 PRR and those that were obtained during the presidential
elections of 1998 and 2000. Clearly, we cannot expect to arrive at any
conclusive results, but these comparisons may help explain how, if at
all, electoral results were altered in 2004.

%f7 ###
\begin{figure*}

\includegraphics{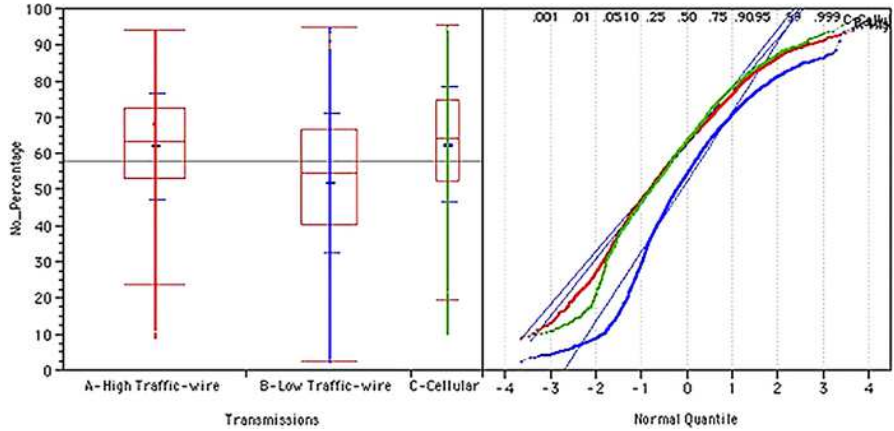}

\caption{Comparison of distributions of  NO\_Percentage
per machine across High, Low Traffic and Cellular groups through box
plots. The  short horizontal straight lines indicate the position of
the means of distributions and standard deviations, the long horizontal
straight lines indicate the position of the percentiles (10\%, 25\%,
median, 75\% and 90\%) of the distribution, and the boxes width shows
the relative size of the samples in High, Low Traffic of wire and
Cellular telephony. On the right-hand side Q--Q plots are shown for the
three empirical distributions showing variances (slopes) for each
distribution.}\label{f7}
\end{figure*}

%s2.2.1 ###
\subsubsection{Percentage of NO votes per machine at the national level}\label{s221}

The comparisons of percentiles, means and medians for the empirical
distributions along with the Van der Waerden test for means are
perfor\-med. Also, when two distributions show close enough means or
medians an analysis of variance is included with its $t$-test to look at
the source of differences.

From previous graphs and numerical tables it is deduced that the
empirical distributions for NO\% per machine in groups\vadjust{\goodbreak} A and C are
equivalent as much in the functional form as in their main quantiles, see Figure~\ref{f7}.
An analysis of variance test comparing means shows there is not
statistical difference between groups A and C with $p=0.4008$ ($p
> 0.05$). Their respective means are $62.0384\pm0.1657$ and $62.3028\pm0.2675$.

The B (Low Traffic) distribution has a Mean and Median significantly
different from those of groups A and C; these differences go up to
around 10 points (20\%). These results together with the
irregular distributions of types of machines in municipalities and
parishes aim at considering that High and Low Traffic groups of
machines cannot be considered representative samples of the electoral
universe as expected. But the Cellular group not expected to produce
electoral results similar to either High or Low traffic groups because
of its geographical distribution is not statistically different to the
High Traffic group, coinciding with the fact that both share the same
pattern of transmission.\looseness=1

The classification of these groups by volume of data transmissions
where the High Traffic and Cellular groups share a pattern quite
different to the Low Traffic one looks like having influence into the
percentage of NO votes per machine.

%s2.2.2 ###
\subsubsection{A comparison to presidential elections of\break 1998 and
2000  by voting centers classified in\break groups~A, B and C}\label{s222}

The next percentage of abstention and empirical distributions for
percentage result in various elections for Chavez 1998, Chavez 2000 and
NO 2004 PRR  per voting center are statistically analyzed. Here we aim
to consider the historical electoral evolution of the centers; we want
to know how different were those centers in the past compared to the
2004 event, as well as how different were their vote results among
groups of centers.

In order to be able to compare the 1998 and 2000 elections with the
2004 PRR event, the percentage of Chavez votes in 1998, 2000 and those
of NO in the 2004~PRR are calculated for each voting center and for the
same centers. This procedure is needed since the structure of electoral
tables was different for those elections. Also, there was a drastic
increase of 32.6\% in the number of voters between the 2000 and the
2004 electoral events.

Voting centers are classified as the High Traffic group (A)
representing 42\% of the universe considered in this study, the Low
Traffic group (B) with 36\% and the Cellular group (C) with 22\%. Each
group contains 44\%, 39\% and 17\% of A, B and C types of machines
respectively. Electoral data are taken from the official results
published by the CNE.

In Table \ref{t1} the relations between groups of voting centers are
detailed.

\begin{figure*}

\includegraphics{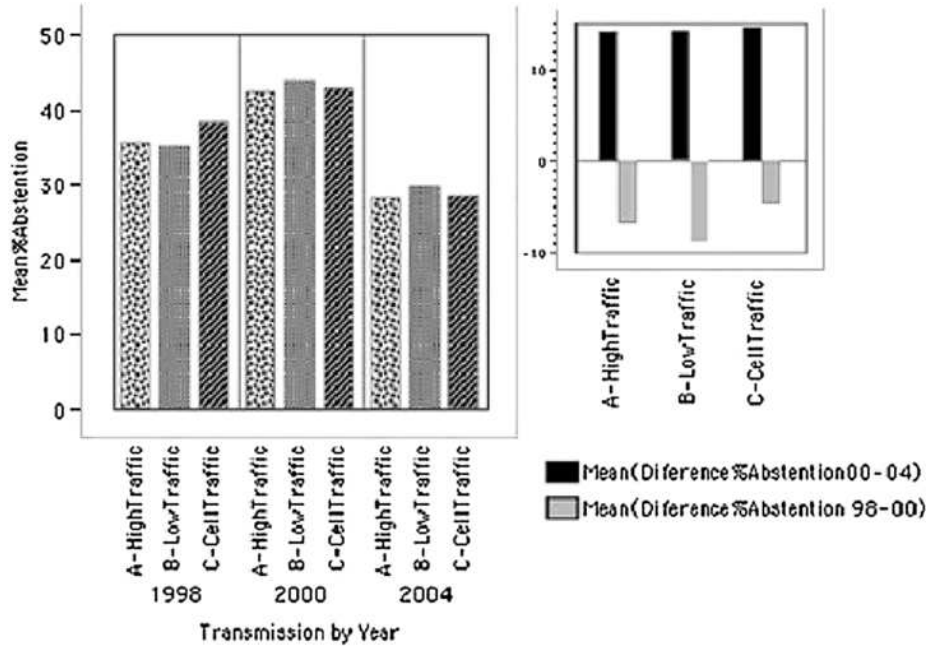}

\caption{Means and differences of \% abstention for
High, Low Traffic and Cellular groups of voting centers in 1998, 2000
and 2004 electoral events.}\label{f8}
\vspace*{-8pt}
\end{figure*}

Comparisons of means and standard deviations of percentage of
abstention in each voting center for the 1998, 2000 and 2004 electoral
events and their differences between successive events for groups A, B
and C are shown in Figure~\ref{f8}.

The number of voting centers analyzed in this section is of 4074: 1759
of them correspond to centers that in 2004 were classified in the High
Traffic group, 1492 in the Low Traffic group and 823 in the Cellular
one.

%f8 ###

In Figure \ref{f8} the most striking finding is that differences in
percentage of abstention by voting center between the 2000 and 2004
electoral events across groups A, B and C are statistically  the same
with $p=0.4275$, when the same measurements between 1998 and 2000 are
different for different groups. In  1998 High and Low Traffic groups
showed means of abstention per center slightly different being lower in
the latter group. As pointed out before, in this group the number of
urban voters is higher; traditionally in Venezuela voters in cities
tend to participate more in elections than rural ones. Group~C behaved
quite  different compared to the others more in line with a rural
behavior. By 2000, the~B group showed a small difference with the other
two groups; abstention increased more in this group than in the others
presumably because there were less Chavez supporters  in it, added to
the fact that the 2000 event was a presidential re-election after a
change of Constitution and just one year and a half of Chavez being
elected with high popularity. But what is unexpected is that the
difference
 with the other groups was maintained in 2004 for a
Presidential Recall Referendum summoned by voters concentrating in
greater numbers precisely in\vadjust{\goodbreak} group~B. It was expected that abstention
would be lower than in\break groups~A and C, even lower than the one
experienced in 1998.

Tables \ref{t2}--\ref{t7} for means, standard deviations and
differences in \% of abstention across groups of voting centers are
shown below.

\begin{figure*}[b]

\includegraphics{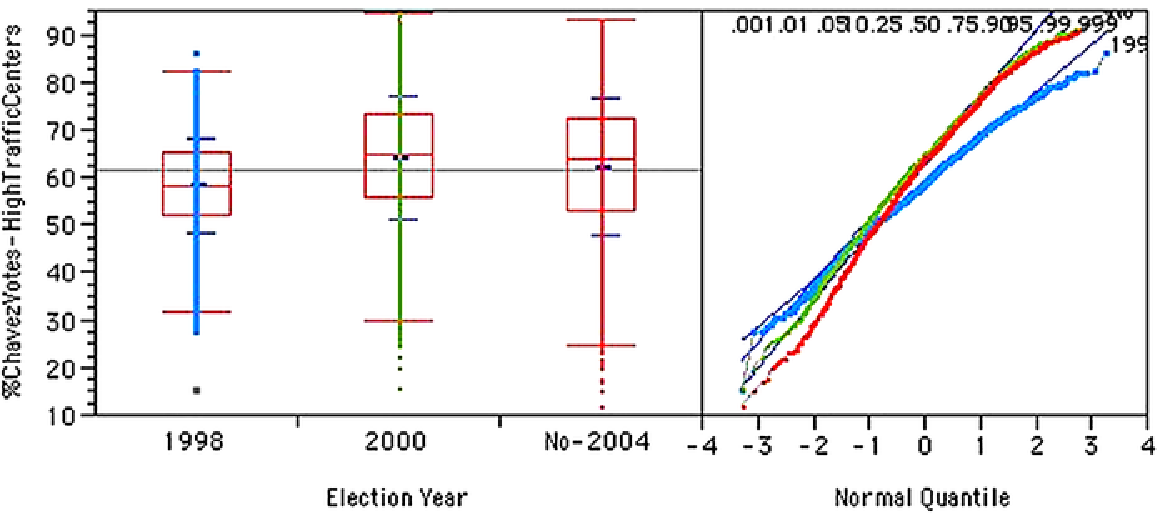}

\caption{Chavez\% votes with respect to valid votes
only for 2004 High Traffic centers (A) in 1998, 2000 and NO\%
PRR.}\label{f9}
\end{figure*}

\begin{figure*}

\includegraphics{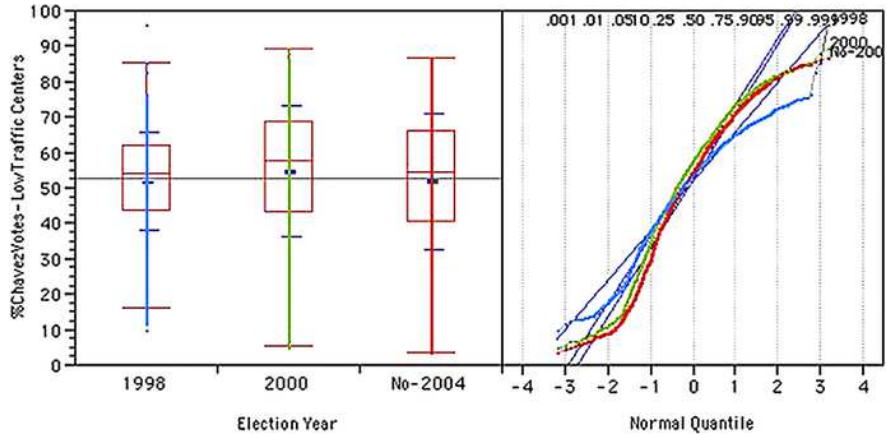}

\caption{Chavez\% votes for 2004 Low Traffic centers (B)
in 1998, 2000 and NO\% PRR.}\label{f10}
\end{figure*}

Comparisons of percentages of Chavez votes distributions  in 1998, 2000
and the 2004 NO votes for the High Traffic and Low Traffic  groups are
shown in Figures~\ref{f9}~and~\ref{f10} as to visualize similarities in
all quantiles between the 2004 PRR and the 2000 presidential election.
Tables with means, standard deviations and quantiles for all groups in
the three electoral events are also included below.

%t2 ###
\begin{table}
\tabcolsep=1pt
\caption{Means, standard deviations and quantiles for percentage of~NO
votes per machine}\label{t2}
{\fontsize{8.35pt}{11.35pt}\selectfont{
\begin{tabular*}{\columnwidth}{@{\extracolsep{\fill}}lcccccc@{}}
\hline
\textbf{Level}&\textbf{Number} &\textbf{Mean}&\textbf{Std dev}&\textbf{25\%---Q}&\textbf{Median}&\textbf{75\%---Q}\\
\hline
% ROW 2
A---High&&&&&&\\
\quad Traffic---wire&8,205&62.0384&14.6481&53.0\phantom{0}&63.51&72.47\\
% ROW 3
B---Low&&&&&&\\
\quad Traffic---wire&7,431&51.8259&19.2504&40.23&54.65&66.59\\
% ROW 4
C---Cellular&3,150&62.3028&15.9224&52.45&63.94&74.54\\
\hline
\end{tabular*}}}
\vspace*{-3pt}
\end{table}

%t3 ###
\begin{table}
\caption{Means, standard deviations of \%  abstention per voting center
for groups A, B and C}\label{t3}
\begin{tabular*}{\columnwidth}{@{\extracolsep{\fill}}lccc@{}}
\hline
% ROW 1
\textbf{Level}&\textbf{Number}&\textbf{Mean}&\textbf{Std dev}\\
\hline
% ROW 2
A---HighTraffic---1998& 1,759& 35.69&6.23\\
% ROW 3
B---LowTraffic---1998& 1,492& 35.05&7.83\\
% ROW 4
C---CellTraffic---1998& \phantom{1,}823& 38.32&8.99\\[4pt]
% ROW 5
A---HighTraffic---2000& 1,759& 42.43&6.70\\
% ROW 6
B---LowTraffic---2000& 1,492& 43.94&8.25\\
% ROW 7
C---CellTraffic---2000& \phantom{1,}823& 42.99&8.63\\[4pt]
% ROW 8
A---HighTraffic---2004& 1,759& 28.35&5.45\\
% ROW 9
B---LowTraffic---2004& 1,492& 29.71&6.34\\
% ROW 10
C---CellTraffic---2004& \phantom{1,}823& 28.41&6.16\\
\hline
\end{tabular*}
\vspace*{-3pt}
\end{table}

From  Figures \ref{f9} and \ref{f10}, it can be perceived that these  groups
show different empirical distributions in their parameters for the
three considered elections. The\vadjust{\goodbreak} distributions of votes percentages in
High Traffic centers are similar for the 2000 and 2004 electoral events
but different from the one in 1998.  The analysis of variance shows
that the means of High Traffic group distributions for the 2000 and the
2004 electoral events are not statistically different with $p=0.0524$
when percentages of Chavez votes are taken with respect to total votes,
that is,  null votes are included. When only valid votes are
considered, means differ but quantiles above median are almost the same
for 2000 and 2004 elections as shown in Figure~\ref{f9}.

%f9 ###

%t4 ###
\begin{table}
\caption{Means, standard deviations for differences in \% abstention
per center between 2000 and 2004 events}\label{t4}\vspace*{-1pt}
\begin{tabular*}{\columnwidth}{@{\extracolsep{\fill}}lccc@{}}
\hline
\textbf{Level}&\textbf{Number}&\textbf{Mean}&\textbf{Std dev}\\
\hline
% ROW 2
A---HighTraffic&1,759&14.09&5.44\\
% ROW 3
B---LowTraffic&1,492&14.23&6.19\\
C---CellTraffic&\phantom{1,}823&14.58&7.12\\
\hline
\end{tabular*}
\vspace*{-9pt}
\end{table}

%t5 ###
\begin{table}
\tabcolsep=0pt
\caption{Means, standard deviations and quantiles for
High~Traffic~centers}\label{t5}\vspace*{-1pt}
\begin{tabular*}{\columnwidth}{@{\extracolsep{\fill}}lcccccc@{}}
\hline
% ROW 1
\textbf{Level}&\textbf{Number}&\textbf{Mean}&\textbf{Std dev}&\textbf{25\%---Q}&\textbf{Median}&\textbf{75\%---Q}\\
\hline
% ROW 2
1998&1,759&58.36&9.90&51.92&58.4\phantom{0}&65.27\\
% ROW 3
2000&1,759&64.11&13.1\phantom{0}&55.76&64.83&73.13\\
% ROW 4
NO---2004&1,759&62.25&14.25&53.15&63.95&72.41\\
\hline
\end{tabular*}
\vspace*{-9pt}
\end{table}

%t6 ###
\begin{table}
\tabcolsep=0pt
\caption{Means, standard deviations and quantiles for Low~Traffic~centers}\label{t6}\vspace*{-1pt}
\begin{tabular*}{\columnwidth}{@{\extracolsep{\fill}}lcccccc@{}}
\hline
% ROW 1
\textbf{Level}&\textbf{Number}&\textbf{Mean}&\textbf{Std dev}&\textbf{25\%---Q}&\textbf{Median}&\textbf{75\%---Q}\\
\hline
% ROW 2
1998&1492&51.62&13.84&43.65&53.94&62.10\\
% ROW 3
2000&1492&54.55&18.53&43.34&57.78&68.71\\
% ROW 4
NO---2004&1492&51.81&19.11&40.61&54.42&66.29\\
\hline
\end{tabular*}
\end{table}

%t7 ###
\begin{table}
\tabcolsep=0pt
\caption{Means, standard deviations and quantiles for Cellular centers}\label{t7}
\begin{tabular*}{\columnwidth}{@{\extracolsep{\fill}}lcccccc@{}}
\hline
% ROW 1
\textbf{Level}&\textbf{Number}&\textbf{Mean}&\textbf{Std dev}&\textbf{25\%---Q}&\textbf{Median}&\textbf{75\%---Q}\\
\hline
1998&823&51.46&11.80&43.36&51.00&60.19\\
% ROW 3
2000&827&60.39&13.38&52.20&60.62&69.80\\
% ROW 4
NO---2004&827&61.80&15.39&52.57&63.33&73.69\\
\hline
\end{tabular*}
\end{table}

There are differences in means in centers classified in the Low Traffic
group for the 2000--2004
years. But, comparing\vadjust{\goodbreak} quantiles, we could appreciate a nearly constant
shift along the entire distribution, a fact that seems surprising. It
is found that 2004 NO\% PRR in the Low Traffic centers have a~mean
statistically comparable to the mean in the 1998 election with
$p=0.7535$.

Also, it is interesting to observe that the Cellular group and the Low
Traffic group show similar statistical behavior in 1998 but in 2000 and
2004 are quite different. The Cellular group resembles more the High
Traffic group in 2004.\vadjust{\goodbreak}

It is worth noticing that the 1998, 2000 and 2004 elections should be
different from a statistical point of view, since in the first one
every voter had to choose from 5 or more options, the second one from
a~maximum of 4 options and in the Recall Referendum choice was only among
2 options in the automated centers. The differences in the number of
options would have to affect the range of votes percentages obtained,
thus, the smaller the number of options is the greater the range of
votes percentages would be. Although this is observed for the ranges of
percentages obtained in the 4074 centers in the 1998, 2000 and 2004
elections (76.42 points), (86.73 points) and (91.08), respectively,
when the mentioned centers are classified into the A, B and C groups,
group~B shows a contraction in the range if the 2000 and the 2004
electoral events are compared. Also, if the standard deviations for
each group  are historically compared, an increase  is observed from
1998 to the 2004 event, nevertheless, the increase from the 2000 to the
2004 PRR is significantly smaller in group~B.

%f10 ###

%t8 ###
\begin{table*}[b]
\caption{}\label{t8}\vspace*{-12pt}
\begin{tabular*}{\textwidth}{@{\extracolsep{4in minus 4in}}l@{}d{4.2}d{1.7}@{\hspace*{2pt}}cd{1.7}@{}d{4.2}d{1.7}@{}}
\hline
&\multicolumn{2}{@{}c@{}}{\textbf{High Traffic centers DF 1758}} & \multicolumn{2}{@{}c@{}}{\textbf{Low Traffic centers DF 1491}}&\multicolumn{2}{@{}c@{}}{\textbf{Cellular centers DF
832}}\\
\ccline{2-3,4-5,6-7}
%&\multicolumn{2}{c}{\textbf{\hrulefill}} & \multicolumn{2}{c}{\textbf{\hrulefill}}&\multicolumn{2}{c@{}}{{\hrulefill}}\\
% ROW 3
\textbf{Distributions}&\multicolumn{1}{c}{\textbf{ChiSq}}&\multicolumn{1}{c}{$\bolds{p}$}&\multicolumn{1}{c}{\textbf{ChiSq}}&\multicolumn{1}{c}{$\bolds{p}$}&\multicolumn{1}{c}{\textbf{ChiSq}}&\multicolumn{1}{c@{}}{$\bolds{p}$}\\
\hline
\%NO 2004---\%Chav 2000 &&&&&&\\
\quad valid votes&1859.99&0.0447&1748.64&3.7\mathrm{E}{-}06&898.92&0.0532\\
% ROW 5
\%NO 2004---\%Chav 2000  &&&&&&\\
\quad total votes&2257.43&4.7\mathrm{E}{-}15&1587.79&0.0402&1135.43&1.0\mathrm{E}{-}11\\
% ROW 6
\%NO 2004---\%Chav 1998 &&&&&&\\
\quad valid votes&5227.07&0&4552.21&3.1\mathrm{E}{-}306&3583.14&0\\
% ROW 7
\%NO 2004---\%Chav 1998 &&&&&&\\
\quad total votes&7810.74&0&5413.84&0&6469.92&0\\
% ROW 8
\%NO 2000---\%Chav 1998 &&&&&&\\
\quad valid votes&3405.52&6.6\mathrm{E}{-}108&3235.13&1.4\mathrm{E}{-}130&3141.21&2.7\mathrm{E}{-}264\\
% ROW 9
\%NO 2000---\%Chav 1998 &&&&&&\\
\quad total votes&4167.38&1.9\mathrm{E}{-}196&3810.55&1.2\mathrm{E}{-}202&4450.29&0\\
\hline
\end{tabular*}
\end{table*}

%s2.2.3 ###
\subsubsection{Chi test of comparison of empirical distributions of Chavez'
votes percentages for 1998, 2000 elections and 2004 PRR}\label{s223}

As it becomes clear in examining the graphs above, there are
similarities between the 2004 PRR and the 2000 elections for automated
voting centers in all groups. So, it proceeds to apply statistical
tests for comparison of the entire distributions. In this case, the Chi
test is used to examine the degree of dependence of data between the
2004 NO\% empirical distribution and the 2000 Chavez\% distribution. By
comparison and reasons of completeness, the Chavez\% vote in the 1998
elections for the same voting centers are included as well. Results are
shown in Table \ref{t8}.

Comparisons are made between the empirical percentage distributions in
the case when only total valid votes are considered and also, when null
votes are included. It occurs for the 1998 and 2000 elections, the 2004
PRR did not have the null option.

It is remarkable that although there was 1 year and 7 months of
difference between the 1998 and 2000 electoral events, Chi test of
comparison between the corresponding empirical distributions show that
they were completely independent events. Nevertheless, the \%NO in 2004
PRR cannot be considered totally independent of the \%Chavez votes in
the 2000 elections  for the High Traffic group ($p=0.0447$), although the
first one was an election of two options and the second one had four
options. There exist differences in the Low Traffic group when
percentages are computed using valid votes only, but when null votes
are included similarities ($p=0.0402$) resemble those of the High Traffic
and Cellular groups ($p=0.0532$).

Something to notice is that the population of voters increased
significantly, 32.6\%, for the 2004 electoral event, but those new
voters do not have to behave from the electoral point of view in the
same manner as the population of the 2000 election.  There are more
similarities between the NO\% 2004 PRR High Traffic and Cellular groups
with the \%Chavez 2000 distributions than with the NO\% 2004 PRR Low
Traffic one. This is unexpected when considering that most of the new
voters are in the former groups. Is this indicating a virtual copying
of 2000 results in the 2004 PRR event or is just a mere coincidence?

%s3 ###
\section{Conclusions}\label{s3}

The programming of electronic voting machines for data transmission or
the programming in the CNE totalizing servers to handle data
transmissions appears to have been different in two groups of machines.
This difference allowed a classification of machines into High Traffic
and Cellular machines with one particular pattern of transmissions, and
Low Traffic machines with quite a different pattern. Differences in the
\mbox{patterns} of transmission across groups cannot be satisfactorily
explained under the electoral rules and technological platforms used.
In fact, they point to two \mbox{different} programs being used either in the
voting machines, totalizing servers or both. The presence of a linear
dependence of transmitted data bytes on votes in both directions in
communications between servers and machines suggests that individual
votes were interchanged in one group of machines. Nonrandomness in the
geographic distribution of groups A, B and C of machines may be showing
intentionality in the differentiation, separating municipalities that
showed higher concentrations of President Chavez supporters in the 2000
election from the rest. Voting machines in these districts were
administered differently than machines in the rest of electoral
districts.

We argue that the percentage of NO votes per machine, as well as the
percentage of abstentions, exhibit a similar distribution across voting
machines in the High Traffic (A) and Cellular (C) groups; the
distribution of both variables is rather different, however, when we
consider machines in the Low Traffic (B) group. The differences in mean
percentage of NO votes and in the percentage of abstentions in machines
of group~B compared to machines of groups A and C are statistically
significant.

The differences in abstention percentages at the center level across
the A, B and C groups for the 1998, 2000 and 2004 electoral events
support the hypothesis of a nonrandom grouping of centers. When
combined with the fact that voting centers of types A and B tended to
be located in different nonoverlapping parishes within the same
municipalities, this may be taken as an indication that tampering in
selected voting centers and selected voting machines may have taken
place.

If indeed tampering occurred, an interesting question is whether
the
2000 election results may have\vadjust{\goodbreak} been approximately reproduced in 2004 to
produce a plausible distribution of NO and YES votes in various
centers. When we compare the distributions of each type of vote across
the elections of 1998, 2000 and 2004 we find that the hypothesis of a
linear dependence between results observed in 2000 and those reported
for 2004 cannot be rejected. We observe a constant shift of the
relative differences of abstentions in centers classified as A, B or C
between 2000 and 2004, an unexpected finding.

While we believe that we have put forth persuasive arguments to
question the integrity of the voting process during the 2004 PRR, our
analyses and conclusions are limited by the fact that voting machines
were not calibrated prior to the election. Thus, even though we cannot
think of a plausible reason for the differences that were observed in
transmission volumes, it is possible that factors totally unrelated to
the electoral process may have had an effect on transmission volumes.
For the monitoring and auditing system of electronic voting machines to
be fully defensible, it would be necessary to calibrate the machines
ahead of the event, perhaps by transmitting a test file of known size
from randomly chosen machines to randomly chosen servers repeatedly so
that the number of bytes used in the transmission can be compared to
the file size.

Deciding whether tampering occurred given the evidence is akin to
deciding between two competing hypothesis: tampering occurred or
tampering did not occur. This decision problem can be formulated as a
posterior odds problem, where we weigh the probability of tampering
given the evidence against the probability of no tampering given the
same evidence. The latter can be thought of the probability of a
coincidental outcome that occurs for reasons which have nothing to do
with tampering. To compute a posterior odds ratio, we need to be able
to evaluate the probability of observing the electoral results (and the
rest of the evidence) we observed under the two hypothesis of tampering
and no tampering. With the information available to us, we can think of
quantifying the conditional probability of the evidence given
tampering. But in order to also quantify the probability of observing
what we observed if no tampering had occurred, we need information that
is not available and that can be obtained through a careful calibration
of the voting machines.

Finally, it is worth mentioning that after the 2006 elections that took
place in Venezuela, the governing party greatly limited the type
and\vadjust{\goodbreak}
amount of information that would be made available about transmissions
between voting machines and the CNE ser\-vers. For example, information
that was available for earlier elections including log headers for
outgoing and incoming data bytes were missing from the transmission
logs  shared with the public. Further, it is no longer possible to
determine the geographic location of each voting machine. Thus, the
analyses that we were able to carry out using the 2004 election data
cannot be carried out for the 2006 election.\looseness=1

\section*{Acknowledgments}

I would like to thank the previous study of telecommunications and
electronic voting technology used during the 2004 PRR made by Ing.
Horacio Velasco and his team. Also, special thanks to Professor Freddy
Malpica who coordinated the investigation group, rendering great
insight into solutions of the puzzle presented. Last, I am grateful to
Professor Lelys Bravo who gave important recommendations as to the
presentation of statistical data and its analysis in the present
article. Further thanks go to Professor Hugo Montesinos for long
clarifying discussions about the most adequate statistical analysis for
the present study.


\begin{thebibliography}{99}

%b1 ###
\bibitem{1}
\textsc{Malpica}, F., \textsc{Velasco}, H. and \textsc{Martin}, I.
Electronic voting in Venezuela: A case study report on the
2004 PRR. Available at
\texttt{\href{http://esdata.info/pdf/VotoElectronico\_es.pdf}{http://esdata.info/pdf/}
\href{http://esdata.info/pdf/VotoElectronico\_es.pdf}{VotoElectronico\_es.pdf}}
(in Spanish),
\texttt{%
\href{http://esdata.info/pdf/ElectronicVote\_en.pdf}{http://}
\href{http://esdata.info/pdf/ElectronicVote\_en.pdf}{esdata.info/pdf/ElectronicVote\_en.pdf}}
(in English) and
\texttt{\href{http://esdata.info/static/video\_transmisiones}{http://esdata.info/static/video\_}
\href{http://esdata.info/static/video\_transmisiones}{transmisiones}}
(in Spanish).

%b2 ###
\bibitem{2}
Log of Transmissions via wire communications network. Cantv
2004 PRR.

%b3 ###
\bibitem{3}
Log of Transmissions via cellular communications network.
Movilnet 2004 PRR.

%b4 ###
\bibitem{4}
Official Results of 2004 PRR by electronic voting machines
published by National Electoral Council CNE. August 2004.
\end{thebibliography}
\end{document}